\documentclass[prb,twocolumn,showpacs,preprintnumbers,amsmath,amssymb,floatfix]{revtex4}

\usepackage{graphicx}
\usepackage{bm}

\begin{document}

\title{Flow boundary conditions for chain-end adsorbing polymer blends}

\author{Xin Zhou}
\author{Denis Andrienko}
\author{Luigi Delle Site}
\author{Kurt Kremer}

\affiliation{Max-Planck-Institut f\"{u}r Polymerforschung,
   Ackermannweg 10, 55128 Mainz, Germany}

\begin{abstract}
  Using the phenol-terminated polycarbonate blend as an example, we
  demonstrate that the hydrodynamic boundary conditions for a flow of
  an adsorbing polymer melt are extremely sensitive to the structure
  of the epitaxial layer. Under shear, the adsorbed parts (chain ends)
  of the polymer melt move along the equipotential lines of the
  surface potential whereas the adsorbed additives serve as the
  surface defects.  In response to the increase of the number of the
  adsorbed additives the surface layer becomes thinner and solidifies.
  This results in a gradual transition from the slip to the no-slip
  boundary condition for the melt flow, with a non-monotonic
  dependence of the slip length on the surface concentration of the
  adsorbed ends.
\end{abstract}

\date{\today}



\maketitle

\section{Introduction}
\label{sec:intro}

The equations of continuum fluid mechanics are incomplete without
appropriate boundary conditions. In most situations it is required
that both normal and tangential components of the relative fluid
velocity vanish at the surface.~\cite{batchelor2000}  This, so called
stick or no-slip boundary condition, has successfully accounted for
most of the experimental facts. It is, however, empirical by nature:
there are no theoretical arguments in favor of the no-slip; moreover,
it has been known since Maxwell's times~\cite{Maxwell67} that even a
simple kinetic theory of gasses predicts the non-zero value of the
tangential velocity at the wall.

Providing some of the insight into the question, kinetic theory fails
already for simple fluids adjacent to a rigid solid. In a more general
context, multiple scattering from the individual wall molecules
remains the major problem of most analytical theories. In this
situation computer simulation techniques are able to advance our
knowledge of the processes occurring at the surfaces.

Indeed, molecular dynamics (MD) simulations demonstrated that, in case
of simple fluids, the flow boundary conditions are sensitive to the
fluid epitaxial order~\cite{Thompson90} as well as the wall
structure.~\cite{cottinbizonne2003} The situation, however, becomes
much more complicated for polymers at the surfaces, because of the
much richer molecular arrangement (e.~g. formation of
brushes,~\cite{klein1996} various adsorbed and depleted
layers~\cite{aubouy1996}) as well as much stronger correlation between
the atoms right at the wall with the rest of the surface
layer.~\cite{ThompsonGR92,harmandaris2005,priezjev2004}

Recent MD studies of the end-adsorbing polymer melts, performed with a
novel quantum based multiscale approach for the surface/polymer
interaction, showed that at least two mechanisms contribute to the
hydrodynamic boundary conditions.~\cite{zhou2005a} The attached parts
of the chains scatter on the surface potential while moving along its
equipotential lines.
This induces the density and the chain conformation modulation in the
adsorbed layer, and energy is lost from these modulations through the
coupling to the thermostat, similar to the situation observed for a
generic model of adsorbed surface layers.~\cite{cieplak1994}
On the other hand, single-end grafted chains of polymer brushes
undergo a coil/stretch transition and disentangle from the melt at a
given shear
rate,~\cite{brochard1992,ajdari1994,smith2005,brochardwyart1994}
favoring the slip boundary condition.
Though investigated for the special case of end adsorbing
polycarbonate on nickel, a similar scenario will occur for block
copolymers with an adsorbing block.

Valid for {\em monodispersed} melts, this picture does not account for
the usual melt polydispersity, which is an outcome of all synthetic
polymerization reactions. Moreover, in many situations the
self-blending of polymer melts is used.~\cite{CheahC03} A small amount
of a lower weight polymer improves the melt processability without
significantly affecting its mechanical properties. It, however, alters
the structure of the surface layer~\cite{andrienko2005a} modifying the
hydrodynamic boundary conditions for the melt flow.  These changes
shall be taken into account at the later stages of melt
processing.~\cite{NamhataGA99}

In this work we focus on the hydrodynamic boundary conditions for
polymer {\em blends} adsorbed on a solid substrate. For this purpose
we consider a particular system, that is bisphenol-A polycarbonate
(BPA-PC) melt sheared over a (111) nickel surface.  Our choice of
polycarbonate as a test system is twofold. First, it is widely used in
various applications~\cite{derudder2000} and, therefore, has been
intensively studied both experimentally~\cite{Morbitzer88} and
theoretically.~\cite{TsaiLC98,TschopKBBH98,TschopKHBB98,DelleSiteAAK02,DelleSiteLK04}
In particular, shear of monodispersed melts~\cite{zhou2005a} and the
structure of the static epitaxial layer for monodispersed
melts~\cite{AbramsK03,AbramsDK03} and blends~\cite{andrienko2005a}
have already been considered. Second, although it is a specific
system, it has a number of important generic features, which also
apply to a realistic description of many polymer brushes composed of
block copolymers.

Our prime goal is to relate the structure of the adsorbed layer, which
changes in the presence of the low molecular weight additive, to the
hydrodynamic boundary conditions for the melt flow, specified by the
slip length and the surface friction coefficient.

\section{Simulation details}

\begin{figure}[ht]
\begin{center}
  \includegraphics[width=6cm]{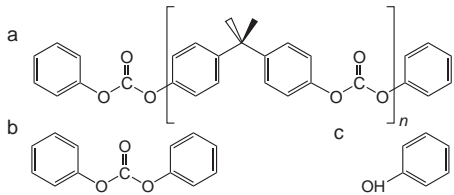}
\end{center}
\caption[]{ 
Chemical structures: (a)
  phenol-terminated bisphenol-{\em A} polycarbonate (in this study
  $n=1,5,20$); (b) diphenyl carbonate (DPC);
  (c) phenol.
}  \label{fig:systems}
\end{figure}

We consider four types of polymer mixtures. The host polymer (major
component) is the phenol terminated BPA-PC of $N_1 = 20$ repeat units.
The second (minor) component is one of the following: BPA-PC of $N_2 =
5$ repeat units, one repeat unit, diphenyl carbonate (DPC), or phenol,
all shown in Fig.~\ref{fig:systems}.
For all mixtures, we use $n_1 = 400$ chains of the major component;
the number of molecules of the minor component, $n_2$, is then
adjusted to provide (approximately) $5\%$ of the total weight of the
system. Exact numbers are given in Table~\ref{tab:systems}.
%
%

\begin{table}[ht]
\begin{center}
\begin{tabular}{|c|c|c|c|c|}
\hline
$N_1,N_2$ & $n_1,n_2$ & $L_z, \sigma$ \\
\hline
20          & 400           & 80.91  \\
20,\,5      & 400,\,76      & 85.15  \\
20,\,1      & 400,\,239     & 84.97  \\
20,\,DPC    & 400,\,521     & 84.69  \\
20,\,Phenol & 400,\,1186    & 83.79  \\
\hline
\end{tabular}
\end{center}
\caption[]{
  Studied systems. $\sigma = 4.41 \AA$.
} \label{tab:systems}
\end{table}

To simulate the melt, we use the previously developed coarse-graining
model, in which each monomer is replaced by four beads that correspond
to isopropylidene, carbonate, and the two linking phenylenes.
Interaction potentials, bead sizes and coarse-graining procedure are
described in Refs.~\cite{AbramsK03,AbramsDK03}.
The bead-wall interaction potentials are obtained from {\em ab~initio}
density functional calculations. All {\em internal} beads experience
strong increasing repulsion below $3.2\, \AA$, either due to the
nature of the molecular interaction or due to the steric hindrance by
the other beads. Only the chain ends absorb on the wall. The wall-end
interaction potential is expanded in 2D reciprocal lattice space of
(111) nickel surface and has the following form
\begin{eqnarray}
U(x,y,z) = \sum_{i} U_{i}(z) \, f_{i}(x,y),
\label{eq:potential}
\end{eqnarray}
where $i=0,1,2$ corresponds to the reciprocal vectors of different
lengths, $f_{0} = 1$, $f_{1}=\cos({\bar x}-{\bar y})+\cos({\bar
  x}+{\bar y})+\cos 2{\bar y}$, and $f_{2}=\cos({\bar x}-3{\bar
  y})+\cos({\bar x} + 3{\bar y})+ \cos 2 {\bar x}$, where $({\bar
  x},{\bar y}) = \frac{2 \pi}{a}\,(x,\frac{y}{{\sqrt 3}})$,

\begin{eqnarray}
\nonumber
U_{0} &=& \left\{
\begin{array}{lll}
 \frac{5}{3} {\epsilon}_{r}
\left[ \frac{2}{5}
\left( \frac{z_0}{z} \right)^{10} -
\left( \frac{z_0}{z} \right)^4 + \frac{3}{5} \right] - \epsilon_0,
&  z < z_{0}  \\
 \frac{\epsilon_0}{2} \left[
\cos \left(\pi \frac{z_c - z}{z_c-z_0} \right)-1 \right],
&  z_0 \le z < z_c
\end{array}
\right.
\end{eqnarray}
\begin{equation}
\nonumber
U_{1,2} = \left\{
\begin{array}{lll}
  - \epsilon_{1,2}, &  z < z_0  \\
    \frac{\epsilon_{1,2} }{2}
    \left[ \cos \left( \pi \frac{z_c - z}{z_c-z_0} \right)-1 \right],
     &   z_0 \le z < z_c
\end{array}
\right.
\end{equation}
The interaction potential obtained using {\em ab initio} calculations
is well reproduced by the following set of parameters: $\epsilon_{r} =
1.5\,
\rm eV$, $\epsilon_{0}= 0.7\, \rm eV$, $\epsilon_{1} = -7/45\, \rm
eV$, and $\epsilon_{2} = -2/45\, \rm eV$.  For details, see
Refs.~\cite{DelleSiteAAK02,DelleSiteAA03,zhou2005a}.

The melts are confined to a slit pore of thickness $L_z$ with the
walls perpendicular to the $z$ axis. Periodic boundary conditions
are employed in $x$ and $y$ directions.
The $x$ and $y$ box dimensions are set to $L_x = 22.23\, \sigma$ and
$L_y = 21.72\,\sigma$ with $\sigma = 4.41\,\AA$, which corresponds to
a (111) hexagonal lattice of nickel with 39 and 22 unit cells. The
number density of beads is $n = 0.85$, which corresponds to $1.05\,
\rm g/cc$, the experimental density at the processing temperature,
$570\, \rm K$. The units are chosen such that $k_{B} T =1$ with 
$T=570\,\rm K$ and $\sigma$ is unity.~\cite{AbramsK03,AbramsDK03}

Starting configurations are generated by randomly placing the chains
in the simulation box. A short run is then used to remove the
bead-bead overlaps.~\cite{AbramsDK03} The production run is performed
in the $NVT$ ensemble with Langevin thermostat with friction $0.5
\tau^{-1}$, where $\tau$ is the unit of time in the simulations. The
thermostat is switched off in the shear direction. The velocity-Verlet
algorithm with the timestep $\Delta t = 0.005 \tau$ is used to
integrate the equations of motion. After equilibration for about $2
\times 10^5$ $\tau$, the shear is applied by moving the top and bottom
walls in opposite directions at a constant velocity $v_w$, so that the
shear rate is $\dot{\gamma} = 2v_{w}/L_z$.
The wall velocity is the same as in our previous studies of {\em
monodispersed} BPA-PC melts,~\cite{zhou2005a} $v_{w} \tau/\sigma =
0.01$. The corresponding shear rate can be obtained from the time
mapping $1 \tau \approx 25\, \rm ps$, see Ref.~\cite{leon2005} for
details.  Using this mapping we obtain $\gamma \approx 10^7 s^{-1}$.
Taking into account that the average chain length in a BPA-PC melt is
$N \sim 70$ the corresponding chain reptation time $\tau_d \sim
N^{3.4}$ is almost by two orders of magnitude larger than that of the
$N=20$ chains. Equivalently, the shear rate would be reduced to
$10^5\, s^{-1}$, close to the value used for industrial processing of
BPA-PC, $\sim 10^4\, s^{-1}$.  Note that smaller shear rates either
have no significant effect on the adsorbed layer or are difficult to
analyze, due to significant error bars for the velocity profiles.

\section{Results}

\subsection{Blends with different additives}

\begin{figure}[ht]
\begin{center}
\includegraphics[width=8cm]{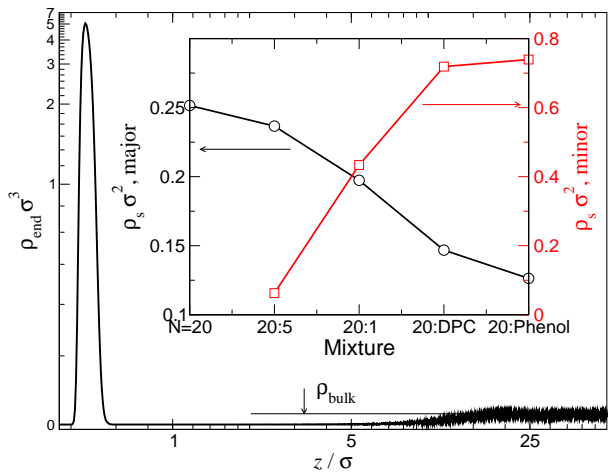}
\end{center}
\caption{
  Typical chain end density profile ($N=20$). The inset depicts the
  change in the surface density (number of the adsorbed chain ends per
  unit area) for the minor and major components. The log scale is used
  to show the bulk density and the depleted region on the same plot.
} \label{fig:end_density}
\end{figure}

We first have a look at the structure of the surface layer.
Fig.~\ref{fig:end_density} shows a typical chain end density profile
for a monodispersed melt ($N=20$). The chain end density has a sharp
peak next to the wall, which is due to the strong adsorption of the
ends; the region with no ends follows; finally, the bulk concentration
is reached at a distance of the order of the radius of gyration, $R_g
\approx 6.2 \sigma = 27\, \AA$.
The inset illustrates how this profile changes in the presence of the
additives: the shorter the chains of the {\em minor} component, the
more of them adsorb on the wall. This of course results in a decrease
of the number of the adsorbed chain ends of the {\em major} component,
in agreement with our earlier studies~\cite{andrienko2005a} except for
the phenol additive, for which we previously observed a weak increase
in the number of the adsorbed chain ends of the major component
compared to the DPC case.~\endnote{The difference is due to the more
  specific, angular-dependent, potential used to describe the end-wall
  attractive interaction in ref.~\cite{andrienko2005a}, which can be
  considered as a refinement of a simpler surface potential used in
  the current study. This affects the conformations of the adsorbed
  molecules at the surface but is not critical for the conclusions of
  the current study: as we will see, the flow boundary conditions are
  basically specified by the surface concentrations of the adsorbed
  chain ends of the major and minor components, and are not very
  sensitive to the conformations of the adsorbed chains.}
%
%

\begin{figure}[ht]
\begin{center}
\includegraphics[width=8cm]{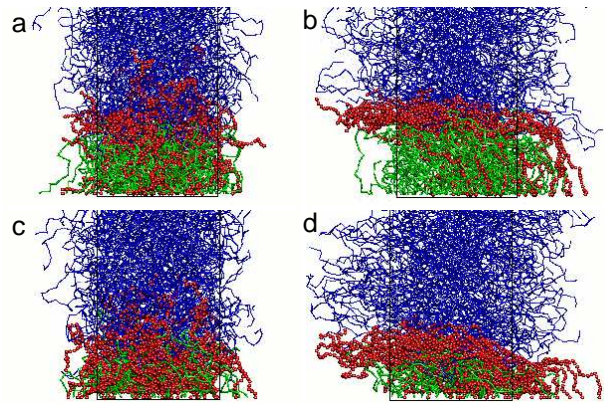}
\end{center}
\caption{ 
 Snapshots of the mixtures: 20:5 without
  (a) and with (b) shear, 20:phenol without (c) and with (d) shear.
  Polymer chains are divided into three populations: chains which
  adsorb both ends (green), only one end (red), and no ends on the
  surface (blue). 
} \label{fig:snapshots}
\end{figure}

The representative snapshots of the systems with and without shear are
shown in Fig.~\ref{fig:snapshots}. Snapshots (a) and (c) illustrate
the thinning of the adsorbed layer due to the decrease in the number
of the adsorbed chain ends of the major component.~\endnote{Note that
  this is not the case for the pure ($N=20$) system and the 20:5
  mixture: the latter has slightly thicker adsorbed layer, which can
  also be seen from the center of mass profiles and the velocity
  profiles. This effect is most probably due to the stiffness of the
  relatively short chains of 5 repeat units, which adsorb and force
  the long chains to stretch away from the wall.}
After shear is applied, the one-end attached chains disentangle from
the melt, stretch and form a thin lubricating layer between the bulk
and the chains adsorbed with two ends (see snapshots (b) and (d)).
Corresponding changes can also be detected by observing the center of
mass profiles.~\cite{zhou2005a}
%
%

\begin{figure}[ht]
\includegraphics[width=8cm]{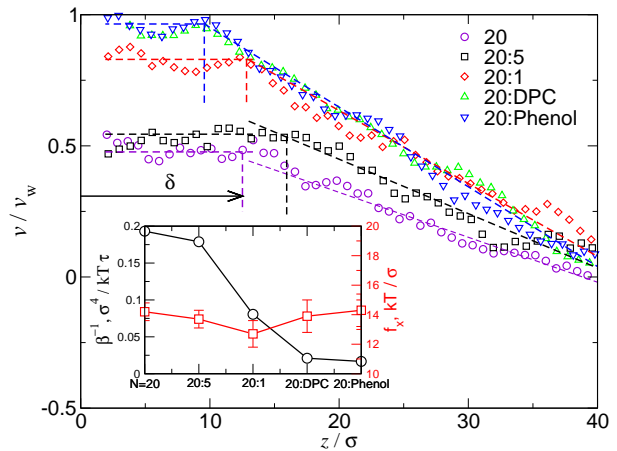}
\caption{
  Velocity profiles normalized to the wall velocity. Only a part of
  the profiles in the vicinity of the wall is shown. The inverse of
  the surface friction coefficient $\beta^{-1} \sim 1 - v_s / v_w$
  (circles) and the shear stress ${f}_x/S$ (squares) are shown in the
  inset. To calculate the velocity profiles we average over all beads,
  independent of the nature of a particular bead.  
} \label{fig:velocity}
\end{figure}

\begin{figure}[ht]
  \includegraphics[width=8cm]{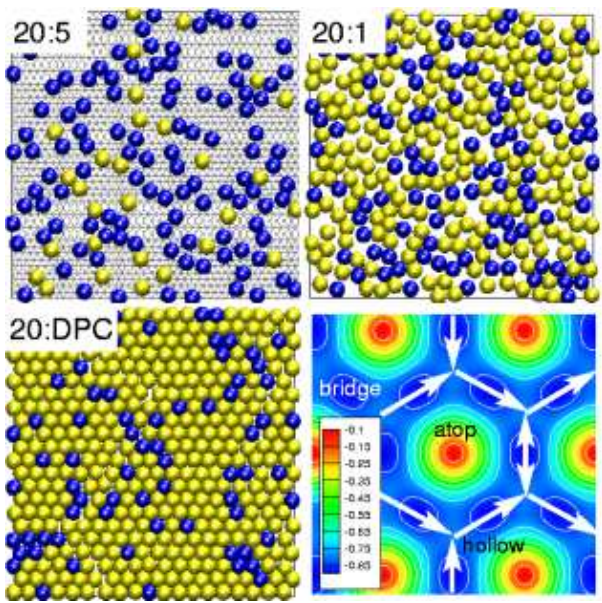}
\caption{ 
  Snapshots of the chain ends adsorbed on the walls. Blue beads: major
  component, yellow beads: minor component. The bead sizes correspond
  to their van der Waals radii. The triangular grid depicts the
  lattice of the (111) nickel surface. Contour plot illustrates the
  bead-wall potential used in simulation, together with the possible
  paths for the adsorbed beads. The contour levels are in electron
  volts, $1\, {\rm eV} \approx 20\, kT$.
} \label{fig:surface}
\end{figure}

Having resolved the structure of the adsorbed layer, we shall turn our
attention to the velocity profiles, which are shown in
Fig.~\ref{fig:velocity}.
For all mixtures, the profiles have features similar to those of
monodispersed systems.~\cite{zhou2005a} Next to the wall the velocity
is practically constant, since the adsorbed chains are dragged by the
wall. Immediately after the plateau, the velocity profile becomes a
linear function of $z$.  The velocity of the beads at the wall, $v_s$,
is smaller than the wall velocity, $v_w$, i.~e. the chain ends slide
over the wall, moving between the hollow and bridge sites of the
potential (the difference in the adsorption energies of these two
sites is rather small, of the order of $2\,kT$, see the contour plot
in Fig.~\ref{fig:surface}).  In case of the monodispersed system and
$20:5$ mixture we have $v_s/v_w \sim 0.5$. This ratio increases for
the $20:1$ mixture and reaches $1$ for the diphenyl carbonate and
phenol additives.
%
%

We can write the stationary shear stress $\tau_{xz}$ (force per unit
area) as~\cite{leger1999}
\begin{eqnarray}
 \tau_{xz} \equiv f_x / S = \beta (v_w - v_s),
\end{eqnarray}
where $\beta$ is the friction coefficient between the adsorbed layer
and the wall, $S$ is the surface area. Since the density of the minor
component in the bulk is small ($5\%$ in weight and most of it is
adsorbed on the walls) the bulk properties of the melt do not change
significantly from mixture to mixture.  Indeed, the shear stress
$\tau_{xz}$ is practically independent of the molecular weight of the
additive (see the inset to Fig.~\ref{fig:velocity}).  Therefore, the
change of $v_s$ is due to the different values of the friction
coefficient $\beta$, which increases significantly for short
additives.

This increase is, in fact, rather unexpected: the adsorbed additives
reduce the number of the attractive sites available for the major
component and, in principle, shall screen the effective interaction of
the melt with the surface. Weaker interaction should result in a
smaller, compared to the monodispersed melt, friction coefficient.
This, however, does not happen in practice: the systems with the
phenol or diphenyl carbonate additives, which basically cover about
80\% of the wall, have the biggest friction coefficient.
%
%

To understand why this happens, let us have a look at the structure of
the beads adsorbed on the surface. Fig.~\ref{fig:surface} shows the
snapshots of the representative mixtures. The difference between the
systems is now clear: the adsorbed layer of the 20:5 mixture has a
structure similar to a two-dimensional (2D) gas; the packing of the
20:1 mixture is more dense, but still disordered, similar to a 2D
liquid; for the 20:DPC as well as 20:Phenol (not shown) mixtures the
concentration of the adsorbed chain ends is so high that they form a
2D crystalline layer. The hexagonal lattice of this layer has the
lattice constant of $2a$, where $a$ is a lattice constant of the (111)
nickel surface, which is of the order of the van de Waals diameter of
the adsorbed beads, i.~e. the close-packed surface layer is not
frustrated energetically, or, in other words, the surface potential
and the surface layer have commensurable lattices.

\subsection{Different concentrations of the phenol additive}

To quantify our results even further, we have also studied mixtures
with different concentrations of the phenol additive. In these
mixtures the surface densities of the adsorbed chain ends and phenol
molecules are monotonic functions of the total number of phenol
molecules in the system. Hence, they can be varied with a good
precision, which helps to analyze the systems in a more systematic
way.

The systems were prepared in a similar manner: the number of the
phenol molecules was adjusted to provide different percentages of the
total weight of the system, in the range from $1\%$ to $20\%$.
Because of the finite size of the simulation box and phenol
adsorption, the concentration of the phenol molecules in the bulk is
always smaller than the one used for the system preparation. We
therefore used this number (from $1\%$ to $20\%$) only to label a
particular system. The important quantities are of course the surface
density of the adsorbed chain ends and phenol molecules.

In addition to the velocity profiles we have calculated the
two-dimensional radial distribution function of the adsorbed molecules
\begin{equation}
g(r) =
\frac{2}{n^2} \frac{L_x L_y}{S(r)} 
\left<
\sum_{i=0}^{n-1} \sum_{j = i+1}^{n} p_{ij}(r)
\right>
\end{equation}
where $r_{ij} = \sqrt{(x_i-x_j)^2 + (y_i-y_j)^2}$, $S(r) =
\pi[(r+\Delta r)^2 - \pi r^2]$ is the bin area, $\Delta r =
r_{max}/N_{bins}$ is the bin width, $p_{ij}(r) = 1$ if
$r<r_{ij}<r+\Delta r$ and zero otherwise. The sum is performed over the
adsorbed beads (chain ends and phenol molecules) only, i.~e.  $z_{i,j}
- z_{wall} < \sigma$; $n$ is the number of the adsorbed beads; $<...>$
denotes the ensemble average.

%
%
\begin{figure}[ht]
\includegraphics[width=8cm]{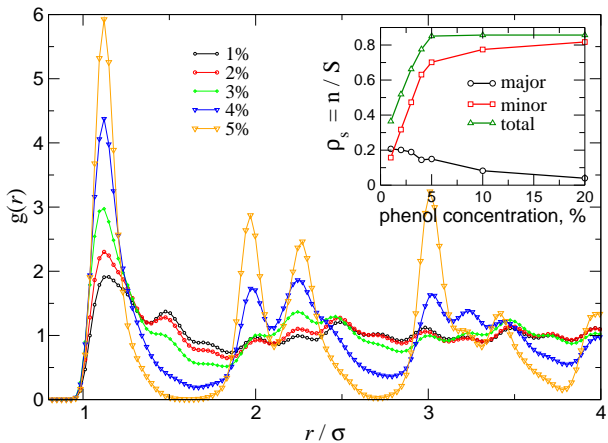}
\caption{
  Two-dimensional radial distribution function of the adsorbed chain
  ends. Inset shows the surface densities (number of adsorbed beads
  per unit area) of the adsorbed ends as a function of the phenol
  concentration.
} \label{fig:rdf}
\end{figure}

To further characterize the quality of the hexagonal packing we have
also calculated the orientational order parameter~\cite{strandberg92}
\begin{equation}
S_6 = \left<
\frac{ 
\left|
\sum_{i=0}^{n-1} \sum_{j = i+1}^{n} q_{ij} \exp(i6\phi_{ij})
\right|
}{
\sum_{i=0}^{n} \sum_{j = i+1}^{n} q_{ij}
}
\right>
\end{equation}
where $\phi_{ij}$ is the angle between the vector $r_{ij}$ and a given
axis in the $xy$ plane, $q_{ij} = 1$ if $r_{ij}$ belongs to the first
peak of the radial distribution function (in our case between $0.5
\sigma$ and $1.5 \sigma$), and zero otherwise. Note that $S_6 = 1$ in
the case of a perfect hexagonal order whereas $S_6 = 0$ indicates the
complete lack of such order.

The radial distribution functions are shown in Fig.~\ref{fig:rdf}. The
increase of the total number of the adsorbed molecules results in a
gradual solidification of the surface layer. The $4\%$ system ($\rho_s
= 0.775 \sigma^{-2}$) has a two-dimensional solid layer formed at the
surface, as it can be seen from the peaks at $2l$ and $\sqrt{3}l$,
where $l$ is the position of the first peak. Further increase of the
amount of adsorbed molecules merely improves the hexagonal
close-packing and, after the $5\%$ system ($\rho_s = 0.851
\sigma^{-2}$) the close packing of the surface is reached, i.~e. the
radial distribution function does not change anymore. The ratio
between the number of the adsorbed chain ends and the adsorbed phenol
molecules, however, still decreases, as it can be seen from the inset
to Fig.~\ref{fig:rdf}. In case of the blends, the solidification of
the surface layer occurs for the 20:DPC mixture.

\begin{figure}[ht]
  \includegraphics[width=8cm]{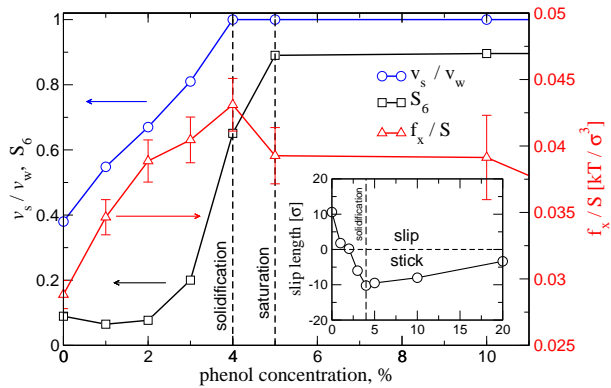}
\caption{ 
  Normalized (to the wall velocity) velocity of the adsorbed layer
  $v_s / v_w$ (circles), orientational order parameter $S_6$
  (squares), and the force per unit area on the wall (triangles).
  Inset: slip length, defined as a distance at which the melt velocity
  extrapolates to zero.
} \label{fig:v_surf}
\end{figure}

The force on the wall and the velocity of the adsorbed layer are shown
in Fig.~\ref{fig:v_surf}, together with the orientational order
parameter $S_6$. The orientational order parameter $S_6$ confirms that
the solidification of the surface layer occurs for the $4\%$ system,
and the saturation for the $5\%$ system, in agreement with the
behavior of the radial distribution functions.

From the dependence of the relative velocity of the adsorbed layer on
the surface density of the adsorbed molecules one can see that once
the epitaxial layer is in a solid state, it follows the wall, i.~e. we
have the stick boundary conditions for the surface layer. However, before
the solidification, the velocity of the adsorbed layer is smaller than
the wall velocity and is roughly proportional to the total
concentration of the adsorbed chain ends.

Similar to the relative velocity, the friction force on the wall
increases with the increase of the total amount of the adsorbed beads.
However, it does not reach a constant value at the point of
solidification of the surface layer, as the relative velocity does,
but starts to decrease, with the decrease rate proportional to the
ratio of the number of the adsorbed chain ends to the number of the
adsorbed phenol molecules. This is due to the fact that the amount of
the {\em long} chains adsorbed on the wall still decreases and the
surface layer disentangles from the bulk of the melt.

Finally, we have also calculated the slip length, which is defined as
the extrapolation length of the linear velocity profile in the bulk to
zero and is often taken as a parameter in macroscopic descriptions of
a fluid flow. The concentration dependence of the slip length is shown
in the inset to Fig.~\ref{fig:v_surf}. Two regimes can be clearly
seen. First, at a low concentration of the adsorbed chain ends (e.~g.
in case of a pure melt, $N=20$) the melt slips over the wall. If we
increase the surface concentration of the adsorbed beads, the slip
length decreases and becomes zero at some concentration, i.~e. we
effectively have the no-slip boundary conditions for the melt flow.
This point, however, does not coincide with the actual locking of the
surface layer, which we observe only when it solidifies.  The reason
can be seen from the shape of the velocity profiles: let us denote the
thickness of the adsorbed layer as $\delta$ and the velocity of this
layer as $v_s$. Then the bulk velocity can be written as $v = v_s
(L_z-2z)/(L_z-2\delta)$. The slip length $b$, obtained from the
condition $v(z = -L_z/2 - b) = v_w$, reads
\begin{equation}
b = \left( v_w / v_s -1 \right) L_z/2 - \delta v_w / v_s.
\label{eq:slip}
\end{equation}
It can be seen that two mechanisms contribute to the total slip length
$b$. The first one is due to the apparent slip of the adsorbed layer
over the surface. The second, negative, contribution is due to the
finite thickness of the adsorbed polymer layer. As we add the
additives, the adsorbed layer shrinks, giving rise to a higher slip;
the relative velocity of the adsorbed layer, however, drops down much
faster.  Combined, these two mechanisms result in a zero slip for some
intermediate concentration of the adsorbed chain ends and a negative
slip for higher surface concentrations.

After the solidification, the adsorbed layer follows the wall, i.~e.
$v_w = v_s$. However, if we further increase the amount of the phenol
additive, the ratio between the number of the adsorbed chain ends of
the long chains and phenol molecules will decrease, as well as the
thickness of the adsorbed layer. As a result, the (negative) slip
length will start to increase again.

%
%

\section{Discussion and conclusions}

Let us first turn to a brief discussion on the possible mechanisms of
friction in our system. Recalling that the adsorbed chain ends move
along the equipotential lines of the surface potential, we conclude
that the role of the adsorbed additives is twofold: apart from
screening the interaction of the melt with the wall, they also serve
as additional obstacles which block the possible paths for the chain
ends of the major component.
If we ignore weak entanglement of the chains in the adsorbed layer as
well as the motion of the adsorbed chain ends of the minor component
over the surface, the situation reduces to a two-dimensional site
percolation problem. Each vertex is designated open or closed at
random, with probability $p$ to be closed which is proportional to the
concentration of the chain ends of the minor component.  Percolation
theory predicts that for the site percolation on a hexagonal lattice
the critical point of the percolation probability is $p_c =
1/2$.~\cite{sykes63} Assuming that the transition happens for the
$20:1$ system (see the inset of Fig.~\ref{fig:velocity}) we obtain
$p_c \approx 0.47$. Of course, due to the applied shear we have an
oriented or directed percolation: our lattice sites shall be assigned
particular orientations, along which the percolative paths are biased.

On the other hand, gradual solidification of the adsorbed layer
results in the increase of the friction coefficient, due to the
collective motion of the adsorbed molecules: once the surface layer is
solidified, the adsorbed molecules cannot move in the direction
perpendicular to the shear direction. However, hopping between the
hole and bridge sites always involves a motion perpendicular to the
shear direction, or, in other words, the hexagonal symmetry of the
lattice forbids the motion of the beads along the shear.

Once the surface layer is in a solid state, the relative velocity of
the adsorbed layer does not change anymore. However, the ratio between
the adsorbed chain ends of long and short molecules still changes.
This affects the thickness of the adsorbed layer as well as its
entanglement with the rest of the melt. As a result, the shear force
on the wall decreases with the increase of the concentration of the
additive. Similar scenario has already been discussed in the framework
of the mean-field
theory.~\cite{brochard1992,brochardwyart1994,ajdari1994,smith2005}

To summarize, we studied the effective boundary conditions for a
polymer blend adsorbed on a structured surface. The slip boundary
condition observed for a monodispersed melt changes to the no-slip at
some concentration of the additive. Further increase of the
concentration of the additive at the surface leads to the
solidification and locking of the motion of the adsorbed surface
layer.

Finally, we would like to comment on the importance of the multiscale
modeling methods we employ in our studies. As underlined in the
introduction, we consider a particular system, which is of high
relevance to different fields of modern technology.  Our results,
despite the particular system considered, are the direct expression of
the interplay between specific (electronic based) molecule-surface
interactions and global statistical and dynamical properties of the
system, i.~e. the interplay between the different scales is the
crucial ingredient of the description. The electronic and molecular
resolution, implicit into the parameterization of the model, allows
for a level of analysis which is beyond any other existing
coarse-grained models or mean-field approaches. In this sense, this
study goes beyond the specificity of the system considered and calls
for extensions to other systems, and experimental tests. The present
approach is a route which allows to detect important, otherwise not
accessible, properties and sets a new link between theory, experiments
and technology.

\acknowledgments This work was supported by the Alexander von Humboldt
foundation, Germany (X.Z.) and by the BMBF, under Grant No. 03N6015,
and the Bayer Corporation. The advise of Vagelis Harmandaris is
acknowledged.

\bibliographystyle{apsrev}
\bibliography{man}

\end{document}